# Presolar stardust in asteroid Ryugu

Jens Barosch[1*], Larry R. Nittler[1], Jianhua Wang[1], Conel M. O'D. Alexander[1], Bradley T. De Gregorio[2], Cécile Engrand[3], Yoko Kebukawa[4], Kazuhide Nagashima[5], Rhonda M. Stroud[2], Hikaru Yabuta[6], Yoshinari Abe[7], Jérôme Aléon[8], Sachiko Amari[9,10], Yuri Amelin[11], Ken-ichi Bajo[12], Laure Bejach[3], Martin Bizzarro[13], Lydie Bonal[14], Audrey Bouvier[15], Richard W. Carlson[1], Marc Chaussidon[16], Byeon-Gak Choi[17], George D. Cody[1], Emmanuel Dartois[18], Nicolas Dauphas[19], Andrew M. Davis[19], Alexandre Dazzi[20], Ariane Deniset-Besseau[20], Tommaso Di Rocco[21], Jean Duprat[8], Wataru Fujiya[22], Ryota Fukai[23], Ikshu Gautam[24], Makiko K. Haba[24], Minako Hashiguchi[25], Yuki Hibiya[26], Hiroshi Hidaka[27], Hisashi Homma[28], Peter Hoppe[29], Gary R. Huss[5], Kiyohiro Ichida[30], Tsuyoshi Iizuka[31], Trevor R. Ireland[32], Akira Ishikawa[24], Motoo Ito[33], Shoichi Itoh[34], Kanami Kamide[6], Noriyuki Kawasaki[12], David A. L. Kilcoyne[35], Noriko T. Kita[36], Kouki Kitajima[36], Thorsten Kleine[37], Shintaro Komatani[30], Mutsumi Komatsu[38,39], Alexander N. Krot[5], Ming-Chang Liu[40], Zita Martins[41], Yuki Masuda[24], Jérémie Mathurin[20], Kevin D. McKeegan[40], Gilles Montagnac[42], Mayu Morita[30], Smail Mostefaoui[8], Kazuko Motomura[43], Frédéric Moynier[16], Izumi Nakai[44], Ann N. Nguyen[45], Takuji Ohigashi[46], Taiga Okumura[31], Morihiko Onose[30], Andreas Pack[21], Changkun Park[47], Laurette Piani[48], Liping Qin[49], Eric Quirico[14], Laurent Remusat[8], Sara S. Russell[50], Naoya Sakamoto[51], Scott A. Sandford[52], Maria Schönbächler[53], Miho Shigenaka[6], Hiroki Suga[54], Lauren Tafla[40], Yoshio Takahashi[31,55], Yasuo Takeichi[55], Yusuke Tamenori[56], Haolan Tang[40], Kentaro Terada[57], Yasuko Terada[54], Tomohiro Usui[23], Maximilien Verdier-Paoletti[8], Sohei Wada[12], Meenakshi Wadhwa[58], Daisuke Wakabayashi[55], Richard J. Walker[59], Katsuyuki Yamashita[60], Shohei Yamashita[55], Qing-Zhu Yin[61], Tetsuya Yokoyama[24], Shigekazu Yoneda[62], Edward D. Young[40], Hiroharu Yui[63], Ai-Cheng Zhang[64], Masanao Abe[23], Akiko Miyazaki[23], Aiko Nakato[23], Satoru Nakazawa[23], Masahiro Nishimura[23], Tatsuaki Okada[23], Takanao Saiki[23], Satoshi Tanaka[23], Fuyuto Terui[65], Yuichi Tsuda[23], Sei-ichiro Watanabe[27], Toru Yada[23], Kasumi Yogata[23], Makoto Yoshikawa[23], Tomoki Nakamura[66], Hiroshi Naraoka[67], Takaaki Noguchi[34], Ryuji Okazaki[67], Kanako Sakamoto[23], Shogo Tachibana[31], Hisayoshi Yurimoto[12,51]

**Affiliations:**

[1]Earth and Planets Laboratory, Carnegie Institution of Washington; 5241 Broad Branch Rd. NW, Washington, DC 20015, USA.

[2]Materials Science and Technology Division, US Naval Research Laboratory; Washington, DC 20375, USA.

[3]IJCLab, UMR 9012, Université Paris-Saclay, CNRS; 91405 Orsay, France.

[4]Faculty of Engineering, Yokohama National University; Yokohama, Kanagawa, 240-8501, Japan.

[5]Hawai'i Institute of Geophysics and Planetology, University of Hawai'i at Mānoa; Honolulu, HI 96822, USA.

[6]Department of Earth and Planetary Systems Science, Hiroshima University; Higashi-Hiroshima, Hiroshima, 739-8526, Japan.

[7]Graduate School of Engineering Materials Science and Engineering, Tokyo Denki University; Tokyo 120-8551, Japan.

[8]Institut de Mineralogie, Physique des Materiaux et de Cosmochimie, Museum National d'Histoire Naturelle, CNRS, Sorbonne Université; Paris, 75231, France.

[9]McDonnell Center for the Space Sciences and Physics Department, Washington University; St. Louis, MO 63130, USA.

[10]Geochemical Research Center, The University of Tokyo; Bunkyo-ku, Tokyo, 113-0033, Japan.

[11]Guangzhou Institute of Geochemistry, Chinese Academy of Sciences; Guangzhou, GD 510640, China.

[12]Department of Natural History Sciences, Hokkaido University; Sapporo, 060-0810, Japan.

[13]Centre for Star and Planet Formation, Globe Institute, University of Copenhagen; Copenhagen, K 1350, Denmark.

[14]Institut de Planétologie et d'Astrophysique, Université Grenoble Alpes; Grenoble, 38000, France.

[15]Bayerisches Geoinstitut, Universität Bayreuth; Bayreuth, 95447, Germany.

[16]Institut de Physique du Globe de Paris, Université de Paris, CNRS; Paris, 75005, France.

[17]Department of Physics and Astronomy, Seoul National University; Seoul, 08826, Republic of Korea.

[18]Institut des Sciences Moléculaires d'Orsay (ISMO), UMR 8214, Université Paris-Saclay, CNRS; Orsay, 91405, France.

[19]Department of the Geophysical Sciences and Enrico Fermi Institute, The University of Chicago; Chicago, IL 60637, USA.

[20]Institut Chimie Physique (ICP), UMR 8000, Université Paris-Saclay, CNRS; Orsay, 91405, France.

[21]Faculty of Geosciences and Geography, University of Göttingen; Göttingen, 37077, Germany.

[22]Faculty of Science, Ibaraki University; Mito, 310-8512, Japan.

[23]Institute of Space and Astronautical Science, Japan Aerospace Exploration Agency; Sagamihara, 252-5210, Japan.

[24]Department of Earth and Planetary Sciences, Tokyo Institute of Technology; Tokyo, 152-8551, Japan.

[25]Graduate School of Environmental Studies, Nagoya University; Chikusa-ku, Nagoya, 464-8601, Japan.

[26]Department of General Systems Studies, The University of Tokyo; Tokyo, 153-0041, Japan.

[27]Department of Earth and Planetary Sciences, Nagoya University; Nagoya, 464-8601, Japan.

[28]Osaka Application Laboratory, Rigaku Corporation; Osaka, 569-1146, Japan.

[29]Max Planck Institute for Chemistry; Mainz, 55128, Germany.

[30]Analytical Technology, Horiba Techno Service Co., Ltd.; Kyoto, 601-8125, Japan.

[31]Department of Earth and Planetary Science, The University of Tokyo; Bunkyo-ku, Tokyo, 113-0033, Japan.

[32]School of Earth and Environmental Sciences, The University of Queensland; St Lucia, QLD 4072, Australia.

[33]Kochi Institute for Core Sample Research, Japan Agency for Marine-Earth Science and Technology; Kochi, 783-8502, Japan.

[34]Division of Earth and Planetary Sciences, Kyoto University; Kitashirakawa Oiwakecho, Sakyo-ku, Kyoto, 606-8502, Japan.

[35]Advanced Light Source, Lawrence Berkeley National Laboratory; Berkeley, CA 94720-8229, USA.

[36]Department of Geoscience, University of Wisconsin-Madison; Madison, WI 53706, USA.

[37]Max Planck Institute for Solar System Research; Göttingen, 37077, Germany.

[38]The Graduate University for Advanced Studies, SOKENDAI; Hayama, Kanagawa, 240-0193, Japan.

[39]Department of Earth Sciences, Waseda University; Shinjuku-ku, Tokyo, 169-8050, Japan.

[40]Department of Earth, Planetary, and Space Sciences, University of California, Los Angeles; Los Angeles, CA 90095, USA.

[41]Centro de Química Estrutural, Institute of Molecular Sciences and Department of Chemical Engineering, Instituto Superior Técnico, Universidade de Lisboa; Av. Rovisco Pais 1, 1049-001, Lisboa, Portugal.

[42]École normale supérieure de Lyon, University Lyon; Lyon, 69342, France.

[43]Thermal Analysis Division, Rigaku Corporation; Tokyo, 196-8666, Japan.

[44]Department of Applied Chemistry, Tokyo University of Science; Tokyo, 162-8601, Japan.

[45]Astromaterials Research and Exploration Science Division, National Aeronautics and Space Administration Johnson Space Center; Houston, TX 77058, USA.

[46]Institute for Molecular Science, UVSOR Synchrotron Facility; Myodaiji, Okazaki, 444-8585, Japan.

[47]Division of Earth-System Sciences, Korea Polar Research Institute; Incheon, 21990, Republic of Korea.

[48]Centre de Recherches Pétrographiques et Géochimiques, Université de Lorraine, CNRS; Nancy, 54500, France.

[49]School of Earth and Space Sciences, University of Science and Technology of China; Anhui, 230026, China.

[50]Department of Earth Sciences, Natural History Museum; London, SW7 5BD, UK.

[51]Isotope Imaging Laboratory, Creative Research Institution, Hokkaido University; Sapporo, 001-0021, Japan.

[52]NASA Ames Research Center; Moffett Field, CA 94035-1000, USA.

[53]Institute for Geochemistry and Petrology, Department of Earth Sciences, Eidgenössische Technische Hochschule Zürich; Zürich, Switzerland.

[54]Spectroscopy and Imaging Division, Japan Synchrotron Radiation Research Institute (JASRI); Sayo-gun, Hyogo, 679-5198, Japan.

[55]Institute of Materials Structure Science, High Energy Accelerator Research Organization, KEK; Tsukuba, Ibaraki, 305-0801, Japan.

[56]Research and Utilization Division, Japan Synchrotron Radiation Research Institute (JASRI); Sayo-gun, Hyogo, 679-5198, Japan.

[57]Earth and Space Science, Osaka University; Osaka, 560-0043, Japan.

[58]School of Earth and Space Exploration, Arizona State University; Tempe, AZ 85281, USA.

[59]Department of Geology, University of Maryland; College Park, MD 20742, USA.

[60]Graduate School of Natural Science and Technology, Okayama University; Okayama, 700-8530, Japan.

[61]Department of Earth and Planetary Sciences, University of California; Davis, CA 95616, USA.

[62]Science and Engineering, National Museum of Nature and Science; Tsukuba, 305-0005, Japan.

[63]Department of Chemistry, Tokyo University of Science; Tokyo, 162-8601, Japan.

[64]School of Earth Sciences and Engineering, Nanjing University; Nanjing, 210023, China.

[65]Kanagawa Institute of Technology; Atsugi, 243-0292, Japan.

[66]Department of Earth Science, Tohoku University; Sendai, 980-8578, Japan.

[67]Department of Earth and Planetary Sciences, Kyushu University; Fukuoka, 819-0395, Japan.

*corresponding author: jbarosch@carnegiescience.edu

**Abstract**

We have conducted a NanoSIMS-based search for presolar material in samples recently returned from C-type asteroid Ryugu as part of JAXA's Hayabusa2 mission. We report the detection of all major presolar grain types with O- and C-anomalous isotopic compositions typically identified in carbonaceous chondrite meteorites: 1 silicate, 1 oxide, 1 O-anomalous supernova grain of ambiguous phase, 38 SiC, and 16 carbonaceous grains. At least two of the carbonaceous grains are presolar graphites, whereas several grains with moderate C isotopic anomalies are probably organics. The presolar silicate was located in a clast with a less altered lithology than the typical extensively aqueously altered Ryugu matrix. The matrix-normalized presolar grain abundances in Ryugu are $4.8\,^{+4.7}_{-2.6}$ ppm for O-anomalous grains, $25\,^{+6}_{-5}$ ppm for SiC grains and $11\,^{+5}_{-3}$ ppm for carbonaceous grains. Ryugu is isotopically and petrologically similar to carbonaceous Ivuna-type (CI) chondrites. To compare the in situ presolar grain abundances of Ryugu with CI chondrites, we also mapped Ivuna and Orgueil samples and found a total of 15 SiC grains and 6 carbonaceous grains. No O-anomalous grains were detected. The matrix-normalized presolar grain abundances in the CI chondrites are similar to those in Ryugu: $23\,^{+7}_{-6}$ ppm SiC and $9.0\,^{+5.4}_{-3.6}$ ppm carbonaceous grains. Thus, our results provide further evidence in support of the Ryugu-CI connection. They also reveal intriguing hints of small-scale heterogeneities in the Ryugu samples, such as locally distinct degrees of alteration that allowed the preservation of delicate presolar material.

## 1. Introduction

Ancient stardust grains occur as trace components in primitive extraterrestrial materials. These tiny (mostly sub-µm) grains condensed in the outflows or explosions of evolved stars, prior to the formation of the Sun (i.e., "presolar"). They can be distinguished from other solar system materials by their highly anomalous isotopic compositions that reflect nucleosynthetic processes in their parent stars. Studying presolar grains in the laboratory allows unique insights into galactic, stellar, interstellar, and asteroidal evolutionary processes (e.g., Zinner 2014; Nittler & Ciesla 2016). With nanoscale secondary ion mass spectrometry (NanoSIMS), the compositions and abundances of presolar grains in astromaterials can be determined in situ. This is particularly useful to compare the characteristics of presolar dust among various extraterrestrial samples, such as chondritic meteorites, interplanetary dust particles and samples returned and to be returned by spacecraft from comets (e.g., NASA Stardust mission) or asteroids (JAXA Hayabusa, Hayabusa2 and NASA OSIRIS-REx missions).

Most presolar grains are either O-rich (oxides, silicates) or C-rich phases (e.g., SiC, graphite), although rare nitrides have also been identified (Nittler et al. 1995). Silicates, the most common presolar phase, can only be studied in relatively pristine samples as they are easily destroyed by secondary processes on asteroids or on Earth (Floss & Haenecour 2016; Nittler et al. 2021; Barosch et al. 2022a). SiC and oxide grains are much more resilient to aqueous alteration and were even detected in highly altered meteorites such as Ivuna-type (CI) chondrite Orgueil (Hutcheon et al. 1994; Huss & Lewis 1995). In Orgueil, a presolar SiC abundance of $14 - 29\ ^{+6}_{-6}$ ppm was estimated from noble gas analyses (Huss & Lewis 1995) and $34\ ^{+11}_{-8}$ ppm was found by NanoSIMS measurements of acid-resistant organic-rich residues (Davidson et al. 2014). Most presolar oxides in Orgueil were found in acid-resistant residues and abundances are not well-quantified (Hutcheon et al. 1994; Dauphas et al. 2010; Qin et al. 2011; Nittler et al. 2018a; Liu et al. 2022). No in-situ study of presolar grain abundances in CI chondrites has been reported to date.

Based on their isotopic compositions, presolar grains can be classified into different groups that link them to their likely stellar sources. The origins of presolar grains and the interpretation of their isotopic compositions have been extensively discussed in the literature (Zinner 2014; Floss & Haenecour 2016, and references therein; Hoppe et al. 2021). Most grains with O-anomalous isotopic compositions belong to one of four O-isotopic groups and are believed to have formed in the winds of asymptotic giant branch (AGB) stars with different masses and/or metallicities (the majority of Group 1–3 grains), or in supernovae (Group 4 and some Group 1 grains). The classification of SiC grains is based on their C, N, and Si isotopic compositions: They are divided into mainstream, AB, C, N, X, Y and Z grains and linked to AGB stars (mainstream, Y, Z), ejecta of novae (N), supernovae (X, C), or of ambiguous origin (AB). Presolar graphite grains are mainly thought to come from AGB stars and supernovae.

The Hayabusa2 spacecraft recently collected and returned ~5.4 g of material from the near-Earth asteroid (162173) Ryugu (Watanabe et al. 2017; Morata et al. 2020; Tachibana et al. 2022). Ryugu is a C-type (carbonaceous) asteroid with a mineralogical, bulk chemical and isotopic composition that closely resembles CI chondrites (Ito et al. 2022; Nakamura et al. 2022; Yada et al. 2022; Yokoyama et al. 2022). The samples consist of carbonate, magnetite and sulfide grains embedded within a phyllosilicate-rich matrix. This mineralogy indicates that extensive aqueous alteration has occurred during water-rock interaction on the Ryugu parent body. CI chondrites also experienced extensive aqueous alteration but might have been further modified on Earth (Brearley 2014). Thus, the returned Ryugu samples provide a unique opportunity to characterize and ascertain their inventory of preserved presolar grains, to compare their abundances and characteristics to the most similar chondritic samples (CI), and to learn about their origins and history. Furthermore, presolar grain abundances may help to identify small-scale heterogeneities in the Ryugu samples, such as in the degree to which Ryugu was modified on its parent body, or heterogeneities in the distribution of presolar material within and between samples.

## 2. Samples and Methods

We analyzed several samples collected by the Hayabusa2 spacecraft during its two touchdowns on the asteroid Ryugu (sample chambers A and C, respectively), as well as material from CI chondrites Orgueil and Ivuna. The following sample types were investigated: (i) Polished thin sections A0058-C1002 (from chamber A; Fig. 1a), C0002-C1001 (from chamber C; Fig. 1b) and Ivuna HK3 (abbreviated as "A0058-2", "C0002" and "Ivuna" in the following; see Yokoyama et al. 2022 for preparation details). (ii) Small Ryugu (A0108-13, C0109-2) and Orgueil grains. These <1 mm-sized grains were crushed between glass slides. A few dozen ~10–30 µm-sized particles were then extracted with a micromanipulator and pressed into annealed gold foil with quartz windows (Fig. 1c). Grains A0108-11 and C0109-8 were sectioned with an ultramicrotome into several ~250 nm thick slices and placed onto Si wafers (see Yabuta et al. 2022 for preparation details). (iii) Lastly, we received insoluble carbonaceous residues that were prepared by acid treatment of Ryugu grains A0106 and C0107 by Yabuta et al. (2022). The residues were deposited onto diamond windows. All samples were Au-coated.

The thin sections were documented with a scanning electron microscope (SEM; JEOL 6500F) and several fine-grained areas were selected for NanoSIMS analyses (see Fig. 1a). These were systematically searched for O- and C-anomalous grains by rastering contiguous 10×10 µm-sized areas with the CAMECA NanoSIMS 50L ion microprobe at the Carnegie Institution, using a $Cs^+$ primary beam (~100–120 nm, ~0.5 pA). Ion images of $^{12}C^-$, $^{13}C^-$, $^{16}O^-$,

$^{17}O^-$, $^{18}O^-$, $^{28}Si^-$, $^{27}Al^{16}O^-$, and secondary electrons were recorded in multi-collection mode. Each area was analyzed for 25 sequential cycles with a resolution of 256×256 pixels and a 1500 µs counting time per pixel per cycle. The pressed particles and organic residues were primarily analyzed to characterize the microscale isotopic variations of organic matter (Barosch et al. 2022b; Yabuta et al. 2022) and were thus not analyzed for O isotopes. Instead, we used the same primary beam conditions to measure $^{12}C_2^-$, $^{12}C^{13}C^-$, $^{12}C^{14}N^-$, $^{12}C^{15}N^-$, plus $^{16}O^-$, $^{28}Si^-$ and $^{32}S^-$ or $^{24}Mg^{16}O^-$, and secondary electrons. An electron gun was used for some measurements to compensate for sample charging. For pressed particles and organic residues larger than 20 µm, the resolution of the ion maps was increased to 512×512 pixels. The counting time was 1000 µs per pixel per cycle and 40 sequential cycles were recorded. To better characterize C-anomalous grains from surrounding materials, some of them were remeasured for N and $^{28}Si^-$, $^{29}Si^-$, and $^{30}Si^-$ isotopes with a higher pixel resolution than used during the initial mapping.

We used the L'image software for data reduction following the protocol described by Nittler et al. (2018b). The ion images were corrected for a 44 ns dead time, for shifts between image frames, and for effects of quasi-simultaneous arrival (Slodzian et al. 2004; Ogliore et al. 2021). In each map, O, C and Si isotopic ratios were internally normalized to the average composition of each image. N isotopic ratios were normalized to atmospheric N and corrected for instrumental mass fractionation using synthetic SiC-$Si_3N_4$ (assumed to have atmospheric $^{15}N/^{14}N$ ratios). Presolar grain candidate regions of interests (ROIs) were identified in sigma images of $\delta^{17}O$, $\delta^{18}O$ and $\delta^{13}C$, in which every pixel represents the number of standard deviations from the average values (see Table 1 for δ-value calculation). The errors are dominated by low counting statistics and were determined from Poisson statistics (cf. Nittler et al. 2018b), which completely dominate the measurements. ROIs include all contiguous pixels within the full width at half maximum of the anomalous region(s) and were considered presolar grains if their isotopic compositions significantly differed from the average compositions. Based on image simulations by Barosch et al. (2022a), we chose the following significance thresholds: 5σ for O- and C-anomalous grains with diameters <200 nm, 4σ for grains with diameters >200 nm and 3.5σ for C-anomalous grains that were clearly associated with $^{28}$Si in the ion images (Fig. 1d, e). A 120 nm beam broadening correction was applied to grains with sizes below 250 nm (Barosch et al. 2022a).

In the following, we classify C-anomalous presolar grains as SiC grains if $^{28}$Si was detected in the ion image or carbonaceous grains if no $^{28}$Si was detected. Carbonaceous grains could be presolar graphite, organic matter or tiny and extremely $^{13}$C-rich SiC grains for which isotopic dilution has erased the intrinsic Si signal (Nguyen et al. 2007). We attempted to determine the mineralogical compositions of O-anomalous grains with the SEM using an Oxford Instruments energy-dispersive X-ray spectrometer (EDX; 5–10 kV accelerating voltage, 1 nA beam current).

## 3. Results

### *3.1. Presolar grain abundances in Ryugu and CI chondrites*

We detected a total of 3 O-anomalous presolar grains, 38 SiC grains, and 16 C-anomalous carbonaceous grains among all of the Ryugu samples. The search of the Orgueil and Ivuna samples resulted in the identification of 15 SiC grains and 6 carbonaceous grains. No O-anomalous grains were found in Ivuna and no O isotopes were measured in Orgueil. All identified presolar grains are listed in Table 1.

Total areas of ~38,700 μm$^2$ and ~25,300 μm$^2$, and ~46,500 μm$^2$ were analyzed in Ryugu chamber A samples, chamber C samples, and CI chondrites, respectively. However, presolar grain abundances were only determined from the thin-section and pressed-particle data. To calculate matrix-normalized presolar grain abundances (Table 2), the total area covered by presolar grains was divided by the total area analyzed, excluding large fragments, cracks, or holes in the samples. Ryugu chamber A samples contain $1.2\,^{+2.8}_{-1.0}$ ppm O-anomalous grains (1σ uncertainties based on Poisson statistics and the uncertainty limits of Gehrels (1986)), $24\,^{+9}_{-7}$ ppm SiC and $17\,^{+8}_{-5}$ ppm carbonaceous grains. Chamber C samples contain $10\,^{+13}_{-7}$ ppm, $26\,^{+11}_{-8}$ ppm and $4.2\,^{+5.5}_{-2.7}$ ppm, respectively (these abundances include both clasts mentioned below; Fig. 2a). The CI chondrite samples contain $23\,^{+7}_{-6}$ ppm SiC and $9.0\,^{+5.4}_{-3.6}$ ppm carbonaceous grains.

All NanoSIMS maps in Ryugu chamber A and the CI chondrite samples were placed randomly in the fine-grained matrix. In section C0002, about one-third of the NanoSIMS maps were placed in the fine-grained matrix (devoid of presolar grains) and the other two-thirds were placed in two areas with less altered lithologies detected by Kawasaki et al. (2022) (clasts 1+2 in Table 2; Fig. 1b; cf. Section 4). Both O-anomalous grains in section C0002 were found in clast 1. The presolar grain abundances for clast 1 are $25\,^{+33}_{-16}$ ppm O-anomalous grains, $47\,^{+38}_{-23}$ ppm SiC and $7.0\,^{+16.2}_{-5.8}$ ppm carbonaceous grains. Clast 2 contains no O-anomalous grains, $11\,^{+11}_{-6}$ ppm SiC and $8.9\,^{+20.5}_{-7.4}$ ppm carbonaceous grains.

*3.2. Presolar grain compositions*

Two O-anomalous grains (HY2-O-01 and -02; Table 1a) are characterized by enrichments of $^{17}O$ and sub-solar to solar $^{18}O/^{16}O$ ratios (Group 1; Fig. 3a), which are believed to mostly indicate formation in low-mass AGB stars (<3 $M_{\odot}$; Zinner 2014). Grain HY2-O-03 is an $^{18}O$-enriched Group 4 grain (likely supernova origin) with a $^{17}O/^{16}O$ ratio close to solar. Grain HY2-O-01 was identified among fine-grained matrix in section A0058-2 and is probably an oxide. It is associated with Al in the ion image which is suggestive of an oxide grain and silicates are unlikely to survive the extensive aqueous alteration seen in most of the Ryugu samples (Fig. 1f; Yokoyama et al. 2022). Grains HY2-O-02 and -03 are both from the same less altered lithology in clast 1 of section C0002 (cf. Section 4). Based on the presence of Si in the EDX map, and the absence of Al in the EDX map and the ion image, HY2-O-02 appears to be a silicate (Fig. 1g, h). Grain HY2-O-03 has the same $Si^-/O^-$ and $AlO^-/O^-$ ratios as the surrounding material and could also be a silicate, although the grain is very small (~120 nm) and it was not possible to cleanly measure its composition.

The C and N isotopic compositions of presolar C-anomalous grains are compared to literature data in Fig. 3b and Si isotopic compositions are displayed in Fig. 3c. Grains without N and Si isotopic data cannot be classified reliably (Table 1b, c). Most Ryugu SiC grains plot in the region of mainstream grains in Fig. 3b and are linked to AGB stars (Zinner 2014). At least two grains with $^{12}C/^{13}C$ ratios of ~5 and ~8 are most likely AB grains, and at least one Z grain (HY2-C-05) was identified by its location on the Si three-isotope plot (Fig. 3c). In the CI chondrite samples, most SiC grains are probably mainstream grains with $^{12}C/^{13}C$ ranging from 12 to 59, but at least two AB grains with $^{12}C/^{13}C \approx 8$ were identified (Ivuna-C-10 and -16). SiC grain diameters range from <100–450 nm, with an average diameter of ~240 nm.

Moderate C-anomalous isotopic compositions of approximately $\delta^{13}C \approx 300$ to -300 ‰ ($^{12}C/^{13}C \approx 68$–127), and a large diversity in $\delta^{15}N$ are typical for ~1% of the C-rich organic grains that are present ubiquitously in Ryugu (Barosch et al. 2022b; Yabuta et al. 2022). Most carbonaceous grains with similar compositions are probably organics. However, the majority of carbonaceous grains reported here have much more anomalous compositions, i.e., with the lowest $^{12}C/^{13}C$ value at 13 and the highest at 224. Some of these could be SiC grains but at least two grains (HY2-C-27, -48) with very low $Si^-/C^-$ ratios are most likely presolar graphites. The C-anomalous grain sizes range from 110–750 nm with an average size of ~270 nm.

## 4. Discussion

Two presolar SiC grains in Ryugu samples were recently detected by Yabuta et al. (2022) and a presolar graphite grain was detected by Ito et al. (2022). Here, we significantly expand on this work. All major types of presolar grains typically found in situ in carbonaceous chondrites were identified in Ryugu: oxides, silicates, SiC, graphites, and C-anomalous organics (Table 1). The detection of at least one presolar silicate (Fig. 1g, h) in Ryugu was particularly unexpected, as silicates are easily destroyed during aqueous alteration (Floss & Haenecour 2016). Their occurrence is probably restricted to relatively rare clasts with less altered lithologies than typical Ryugu matrix, as described by Kawasaki et al. (2022; clast 1 in Fig. 1b). These contain anhydrous minerals such as Mg-rich olivines and low-Ca pyroxenes, which indicate that these clasts escaped extensive alteration. The inferred abundance of O-anomalous grains in C0002 clast 1 is $25\,^{+33}_{-16}$ ppm (2 grains; 4907 μm$^2$ analyzed; Fig. 2a; no O-anomalous grains were found in clast 2) and consistent with fairly altered CM2 and CR2 chondrites (Leitner et al. 2020, and references therein). Nguyen et al. (2022) reported an abundance of $165\,^{+161}_{-90}$ ppm (3 O-anomalous grains; 1072 μm$^2$ analyzed) for a similar clast in another Ryugu chamber C section. While much more analysis is required to determine reliable abundances, it is clear that these clasts were able to preserve delicate presolar grains, whereas typical Ryugu matrix was not.

No O-anomalous grains were detected in situ in Ivuna, although oxides are known to exist in CI chondrites (Hutcheon et al. 1994). Assuming an average presolar grain size of 0.25 μm, we estimate a 1σ upper limit for non-detection for the abundances of O-anomalous grains of ~4 ppm (Gehrels 1986). This upper limit is fully consistent with the abundances seen in the Ryugu matrix ($1.2\,^{+2.8}_{-1.0}$ ppm) and the initial estimate of ~0.5 ppm by Hutcheon et al. (1994) based on a single large presolar $Al_2O_3$ grain from Orgueil. Recent studies by Morin et al. (2022) and Kawasaki et al. (2022) report Ivuna clasts that contain anhydrous primary minerals likely derived from amoeboid olivine aggregates and/or chondrules. These lithologies are similar to the less altered lithologies in Ryugu and might be interesting future targets for presolar grain analyses.

The inferred abundances of SiC grains are consistent across Ryugu chamber A and C samples ($24\,^{+9}_{-7}$ ppm vs. $26\,^{+11}_{-8}$ ppm), and essentially the same abundances seen in the CI chondrite samples ($23\,^{+7}_{-6}$ ppm). Independent (ex-situ) estimates of the SiC abundance in CI chondrites are similar within uncertainties (Fig. 2a,b; Huss & Lewis 1995; Davidson et al. 2014). This result is thus consistent with the Ryugu-CI chondrite connection (Ito et al. 2022; Nakamura et al. 2022; Yada et al. 2022; Yokoyama et al. 2022); however, comparable SiC abundances are also found in other unheated carbonaceous chondrites (Fig. 2b).

SiC grains were relatively homogeneously distributed across the Ryugu thin sections, with at least one SiC found in most Ryugu matrix regions measured (Fig. 1a). In contrast, their distribution in Ivuna seems much more heterogeneous: 60% of all grains were detected in relatively close proximity in a ~0.5×0.5 mm-sized matrix region which was indistinguishable by SEM from the other regions measured, with a factor of three lower number density in the other regions. The abundances of carbonaceous grains seem to be slightly higher in Ryugu chamber A samples compared to chamber C samples and CI chondrites ($17\,^{+8}_{-5}$ ppm vs. $4.2\,^{+5.5}_{-2.7}$ ppm vs. $9.0\,^{+5.4}_{-3.6}$ ppm, respectively). These differences might indicate some heterogeneity in the distribution of presolar material in the matrices of these samples.

As in most other in-situ studies, the presolar grain abundances reported here are lower limits. Particularly small grains (<100 nm) can only be reliably identified by higher-resolution NanoSIMS analyses (Hoppe et al. 2015; Nittler et al. 2018a). Some presolar grain types found in Orgueil, such as nanodiamonds (Zinner 2014) and presolar oxide grains with anomalies in isotopes other than oxygen (e.g., $^{54}Cr$ and $^{50}Ti$; Dauphas et al. 2010; Qin et al 2011; Nittler et al. 2018a), were missed in this study.

The presolar grains detected in Ryugu have isotopic compositions consistent with those seen in primitive meteorites (Zinner et al. 2014, and references therein). Compared to the isotopic compositions of SiC grains detected in acid residues (cf. presolar grain database; Stephan et al. 2020), the N isotopic ratios for many Ryugu SiC grains are closer to solar and/or terrestrial (Fig. 3b). Moreover, the $^{12}C/^{13}C$ ratios of five Ryugu SiC grains are between 10 and 20 (Table 1b), which is an unusual value, falling between the AB and mainstream SiC populations. Both observations can be explained by the ubiquity of organic matter in Ryugu (Barosch et al. 2022b; Yabuta et al. 2022). Indeed, it was not always possible to completely disentangle the C and N signals arising from organics from those intrinsic to the presolar grains. The contribution of C and N from surrounding organic matter to the grains dilute the isotopic signatures of presolar SiC grains and shift them toward less anomalous compositions, as shown by the green model mixing curves in Fig. 3b. These curves indicate mixing between six select "true" SiC compositions and bulk organic matter in Ryugu ($^{12}C/^{13}C \approx 90$, $^{14}N/^{15}N \approx 260$; Yabuta et al. 2022), and show how C and N contamination from the organic matter leads to a narrower range of SiC compositions compared to literature data. This could also lead to misclassification and/or non-identification of C-anomalous presolar grains with low to moderate anomalies in Ryugu and CI chondrites.

## 5. Conclusions

The samples returned from asteroid Ryugu by the Hayabusa2 spacecraft contain presolar stardust grains. Their abundances and compositions are similar to presolar material found in CI chondrites. Thus, our results provide further evidence that asteroid Ryugu is closely related to CI chondrites, a connection originally based on mineralogical and bulk chemical and isotopic data (Ito et al. 2022; Nakamura et al. 2022; Yada et al. 2022; Yokoyama et al. 2022).

Refractory O- and C-rich presolar phases survived the pervasive aqueous alteration that Ryugu has experienced, whereas delicate presolar silicates were likely destroyed. However, small regions of Ryugu escaped extensive alteration (Kawasaki et al. 2022; Nakamura et al. 2022) and allowed their preservation. Further analyses of less altered Ryugu lithologies would be highly beneficial to better characterize their inventory of preserved presolar material and compare it to more altered Ryugu matrix.

Petrographically and isotopically similar less altered clasts were recently detected in Ivuna (Kawasaki et al. 2022; Morin et al. 2022) and could be targeted in future studies for comparison with the Ryugu samples. These clasts might contain presolar silicates that have not yet been found in Ivuna. The presence or absence of presolar material in these clasts would provide important clues about their origin and their history of secondary processing.

Future NanoSIMS-based analyses of Ryugu samples will focus on particles with shallower 2.7 μm OH absorption features in their infrared reflectance spectra (cf. Yada et al. 2022). These may be less aqueously altered. Their study will allow us to better assess the scale of heterogeneity sampled on Ryugu, and to explore the effects of differing degrees of alteration on organics (cf. Yabuta et al. 2022) and presolar grains. Systematic searches for presolar grains in all Ryugu lithologies will provide a representative dataset of presolar grain abundances and characteristics in asteroid Ryugu and will extract the maximum scientific information from these precious samples.

## Acknowledgements

This work was carried out under the auspices of the Hayabusa2 Initial Analysis Team, specifically the Chemistry sub-team led by Prof. H. Yurimoto and the Macromolecule sub-team led by Prof. H. Yabuta. We thank Nico Küter for assistance during sample preparation. This work was funded in part by NASA Grants NNX16AK72G and 80NSSC20K0340 to LRN.

## References

Barosch, J., Nittler, L. R., Wang, J., et al. 2022a, GeCoA, in review

Barosch, J., Nittler, L. R., De Gregorio B. T., et al. 2022b, 53rd LPSC, abstr.#2050

Brearley, A. J. 2014, In Meteorites and Cosmochemical Processes (ed. A. M. Davis), Vol. 1, Treatise on Geochemistry 2nd ed. (exec. eds. H. D. Holland & K. K. Turekian), 309

Dauphas, N., Remusat, L., Chen, J. H., et al. 2010, ApJ, 720, 1577

Davidson, J., Busemann, H., Nittler, L. R., et al. 2014, GeCoA, 139, 248

Floss, C., & Stadermann, F. J. 2012, M&PS, 47, 992

Floss, C., & Haenecour, P. 2016, GeocJ, 50, 3

Gehrels, N. 1986, ApJ, 303, 336

Haenecour, P., Floss, C., Zega, T. J., et al. 2018, GeCoA, 221, 379

Hoppe, P., Leitner, J., & Kodolányi, J. 2015, ApJ, 808, 9

Hoppe, P., Leitner, J., Kodolányi, J., & Vollmer C. 2021, ApJ, 913, 10

Huss, G. R., & Lewis, R. S. 1995, GeCoA, 59, 115

Hutcheon, I. D., Huss, G. R., Fahey, A. J., & Wasserburg, G. J. 1994, ApJ, 425, L97

Ito, M., Tomioka, N., Uesugi, M., et al. 2022, NatAs, in press

Kawasaki, N., Nagashima, K., Sakamoto, N., et al. 2022, Sci, in review

Leitner, J., Hoppe, P., & Zipfel, J. 2018, M&PS, 53, 204

Leitner, J., Metzler, K., Vollmer, C., et al. 2020, M&PS, 55, 1176

Liu, N., Dauphas, N., Cristallo, S., Palmerini, S., & Busso, M. 2022, GeCoA, 319, 296

Morata, T., Sugita, S., Cho, Y., et al. 2020, Sci, 368, 654

Morin, G. L. F., Marrocchi, Y., Villeneuve, J., & Jacquet, E. 2022, GeCoA, in press

Nakamura, E., Kobayashi, K., Tanaka, R., et al. 2022, Proc. Jpn. Acad. Ser. B, 98, 227

Nguyen, A. N., Stadermann, F. J., Zinner, E., et al. 2007, ApJ, 656, 1223

Nguyen, A. N., Mane, P., & Piani, L. 2022, 85th MetSoc meeting, abstr.#6376

Nittler, L. R., Hoppe, P., Alexander, C. M. O'D., et al. 1995, ApJ, 453, L25

Nittler, L. R., & Ciesla, F. 2016, ARA&A, 54, 53

Nittler, L. R., Alexander, C. M. O'D, Liu, N., & Wang, J. 2018a, ApJL, 856, L24

Nittler, L. R., Alexander, C. M. O'D., Davidson, J., et al. 2018b, GeCoA, 226, 107

Nittler, L. R., Alexander, C. M. O'D., Patzer, A., & Verdier-Paoletti, M. J. 2021, M&PS, 56, 260

Ogliore, R., Nagashima, K., Huss, G., & Haenecour, P. 2021, NIMPA, 491, 17

Qin, L., Nittler, L. R., Alexander, C. M. O'D., et al. 2011, GeCoA 75, 629

Slodzian, G., Hillion, F., Stadermann, F. J., & Zinner, E. 2004, ApSS, 231, 874

Stephan, T., Bose, M., Boujibar, A., et al. 2020, 51st LPSC, abstr. #2140

Tachibana, S., Sawada, H., Okazaki, R., et al. 2022, Sci. 375, 1011

Watanabe, S., Tsuda, Y., Yoshikawa, M., et al. 2017, SSRv, 208, 3

Yabuta, H., Cody, G. C., Engrand, C., et al. 2022, Sci, in review

Yada, T., Abe, M., Okada, T., et al. 2022, NatAs, 6, 214

Yokoyama, T., Nagashima, K., Nakai, I., et al. 2022, Sci, in press

Zhao, X., Floss, C., Lin, Y., & Bose, M. 2013, ApJ, 769, 49

Zinner, E., Amari, S., Guinness, R., et al. 2007, GeCoA, 71, 4786

Zinner, E., 2014, In Meteorites and Cosmochemical Processes (ed. A. M. Davis), Vol. 1, Treatise on Geochemistry 2nd ed. (exec. eds. H. D. Holland & K. K. Turekian), 181

# Tables

## Table 1: Characteristics of O- and C-anomalous grains identified in Ryugu and CI chondrites.

### a) Ryugu O-anomalous grains

| Name | Sample | Type | Chamber | Phase | Group | Size (µm)* | $^{17}O/^{16}O$ (x$10^{-4}$) | $^{18}O/^{16}O$ (x$10^{-3}$) | $Si^-/O^-$ | $AlO^-/O^-$ |
|---|---|---|---|---|---|---|---|---|---|---|
| HY2-O-01 | A0058-2 | thin section | A | oxide | 1 | 0.17 | 20.89 ± 0.95 | 1.84 ± 0.09 | 0.015 | 0.006 |
| HY2-O-02 | C0002 | thin section | C | silicate | 1 | 0.38 | 5.57 ± 0.34 | 1.14 ± 0.06 | 0.017 | 0.003 |
| HY2-O-03 | C0002 | thin section | C | ambiguous | 4 | 0.12 | 3.75 ± 0.61 | 2.92 ± 0.17 | 0.010 | 0.004 |

### b) Ryugu C-anomalous grains

| Name | Sample | Type | Chamber | Phase | Group | Size (µm)* | $\delta^{13}C$ (‰) | $Si^-/C^-$ | $^{12}C/^{13}C$ | $\delta^{15}N$ (‰) | $\delta^{29}Si$ (‰) | $\delta^{30}Si$ (‰) |
|---|---|---|---|---|---|---|---|---|---|---|---|---|
| HY2-C-01** | A0108-11 | mtome | A | SiC | ms or AB | 0.12 | 4000 ± 240 | - | 17.8 | -221 ± 247 | | |
| HY2-C-02** | A0108-11 | mtome | A | C-anom | ambiguous | 0.23 | 360 ± 57 | - | 65.4 | 94 ± 183 | | |
| HY2-C-03 | C0109-8 | mtome | C | C-anom | ambiguous | 0.15 | 1796 ± 162 | - | 31.8 | -353 ± 264 | 22 ± 11 | 16 ± 14 |
| HY2-C-04 | C0109-2 | pressed particle | C | SiC | ms, X or Z | 0.37 | 657 ± 21 | 0.76 | 53.7 | -552 ± 141 | 162 ± 15 | 187 ± 19 |
| HY2-C-05 | C0109-2 | pressed particle | C | SiC | Z | 0.19 | 606 ± 38 | 0.77 | 55.4 | -155 ± 203 | -87 ± 23 | 62 ± 29 |
| HY2-C-06 | C0109-2 | pressed particle | C | SiC | AB | 0.11 | 4770 ± 101 | 0.82 | 15.4 | 705 ± 428 | 228 ± 24 | 195 ± 29 |
| HY2-C-07 | A0108-13 | pressed particle | A | SiC | AB | 0.17 | 10832 ± 149 | 0.87 | 7.5 | -825 ± 189 | -33 ± 21 | -21 ± 26 |
| HY2-C-08 | A0108-13 | pressed particle | A | SiC | ms or AB | 0.14 | 4542 ± 212 | 1.12 | 16.1 | 200 ± 231 | -25 ± 39 | -18 ± 48 |
| HY2-C-09 | A0108-13 | pressed particle | A | SiC | ms, X or Z | 0.14 | 1005 ± 62 | 0.32 | 44.4 | -236 ± 161 | 29 ± 37 | 58 ± 46 |
| HY2-C-10 | A0108-13 | pressed particle | A | SiC | ms, X or Z | 0.19 | 300 ± 35 | 0.92 | 68.5 | -25 ± 101 | 59 ± 16 | 98 ± 19 |
| HY2-C-11 | A0058-2 | thin section | A | SiC | ms or AB | 0.26 | 6070 ± 330 | 0.53 | 12.6 | 114 ± 180 | 24 ± 77 | -82 ± 93 |
| HY2-C-12 | A0058-2 | thin section | A | SiC | ms, X or Z | 0.27 | 1278 ± 126 | 0.39 | 39.1 | -152 ± 49 | 0 ± 8 | 45 ± 11 |
| HY2-C-13 | A0058-2 | thin section | A | SiC | ms, X or Z | 0.29 | 549 ± 48 | 0.73 | 57.4 | -11 ± 64 | -40 ± 23 | -38 ± 29 |
| HY2-C-14 | A0058-2 | thin section | A | C-anom | ambiguous | 0.21 | 805 ± 124 | 0.24 | 49.3 | | | |
| HY2-C-15 | A0058-2 | thin section | A | SiC | ms, X or Z | 0.30 | 2362 ± 185 | 0.56 | 26.5 | 117 ± 178 | -5 ± 42 | 71 ± 53 |
| HY2-C-16 | A0058-2 | thin section | A | SiC | ms, X or Z | 0.29 | 461 ± 74 | 0.60 | 60.9 | | | |
| HY2-C-17 | A0058-2 | thin section | A | SiC | ms, X or Z | 0.25 | 340 ± 56 | 0.62 | 66.4 | -146 ± 86 | 99 ± 20 | 79 ± 24 |
| HY2-C-18 | A0058-2 | thin section | A | C-anom | ambiguous | 0.14 | 1294 ± 250 | 0.43 | 38.8 | | | |
| HY2-C-19 | A0058-2 | thin section | A | C-anom | likely organic | 0.23 | -296 ± 56 | 0.12 | 126.5 | | | |
| HY2-C-20 | A0058-2 | thin section | A | SiC | ms, X or Z | 0.14 | 709 ± 173 | 0.60 | 52.1 | | | |
| HY2-C-21 | A0058-2 | thin section | A | C-anom | ambiguous | 0.11 | 2092 ± 297 | 0.62 | 28.8 | | | |
| HY2-C-22 | A0058-2 | thin section | A | C-anom | likely organic | 0.28 | -243 ± 38 | 0.14 | 117.6 | -79 ± 35 | -6 ± 20 | 34 ± 25 |
| HY2-C-23 | A0058-2 | thin section | A | C-anom | ambiguous | 0.19 | 2603 ± 430 | 0.20 | 24.7 | 55 ± 100 | -9 ± 60 | 31 ± 75 |
| HY2-C-24 | A0058-2 | thin section | A | C-anom | ambiguous | 0.18 | 5885 ± 423 | 0.70 | 12.9 | -222 ± 225 | 28 ± 57 | 14 ± 69 |
| HY2-C-25 | A0058-2 | thin section | A | SiC | ms, X or Z | 0.13 | 942 ± 154 | 0.35 | 45.8 | -144 ± 129 | 72 ± 57 | -4 ± 67 |
| HY2-C-26 | A0058-2 | thin section | A | C-anom | ambiguous | 0.19 | 2116 ± 61 | 1.18 | 28.6 | 83 ± 99 | -16 ± 30 | -29 ± 37 |
| HY2-C-27 | A0058-2 | thin section | A | C-anom | graphite | 0.41 | -604 ± 6 | 0.16 | 224.5 | 18 ± 41 | -7 ± 15 | 14 ± 18 |
| HY2-C-28 | A0058-2 | thin section | A | SiC | ms, X or Z | 0.30 | 507 ± 29 | 5.05 | 59.1 | -70 ± 81 | 35 ± 10 | 57 ± 12 |
| HY2-C-29 | C0002 | thin section | C | SiC | AB | 0.31 | 15360 ± 266 | 0.85 | 5.4 | | | |
| HY2-C-30 | C0002 | thin section | C | SiC | ms, X or Z | 0.28 | 555 ± 71 | 0.73 | 57.2 | | | |
| HY2-C-31 | C0002 | thin section | C | C-anom | ambiguous | 0.21 | 664 ± 124 | 0.31 | 53.5 | | | |
| HY2-C-32 | C0002 | thin section | C | SiC | ms, X or Z | 0.21 | 2217 ± 141 | 0.75 | 27.7 | | | |
| HY2-C-33 | C0002 | thin section | C | SiC | ms or AB | 0.14 | 3917 ± 621 | 0.74 | 18.1 | | | |
| HY2-C-34 | C0002 | thin section | C | SiC | ms, X or Z | 0.14 | 2598 ± 221 | 0.33 | 24.7 | | | |
| HY2-C-35 | C0002 | thin section | C | C-anom | ambiguous | 0.20 | -329 ± 63 | 0.06 | 132.7 | | | |
| HY2-C-36 | C0002 | thin section | C | SiC | ms, X or Z | 0.12 | 1923 ± 318 | 0.33 | 30.4 | | | |
| HY2-C-37 | C0002 | thin section | C | SiC | ms, X or Z | 0.27 | 636 ± 92 | 0.88 | 54.4 | | | |
| HY2-C-38 | A0106-IOM | IOM-residue | A | C-anom | likely organic | 0.51 | 73 ± 10 | 0.00 | 82.9 | 3 ± 29 | | |
| HY2-C-39 | A0106-IOM | IOM-residue | A | C-anom | likely organic | 0.75 | 91 ± 8 | 0.00 | 81.6 | 129 ± 31 | | |
| HY2-C-40 | A0106-IOM | IOM-residue | A | SiC | ms, X or Z | 0.45 | 424 ± 22 | 0.49 | 62.5 | -150 ± 72 | | |
| HY2-C-41 | A0106-IOM | IOM-residue | A | SiC | ms, X or Z | 0.32 | 322 ± 21 | 0.05 | 67.3 | -260 ± 60 | | |
| HY2-C-42 | A0106-IOM | IOM-residue | A | SiC | ms, X or Z | 0.31 | 106 ± 23 | 0.02 | 80.4 | -116 ± 68 | | |
| HY2-C-43 | A0106-IOM | IOM-residue | A | SiC | ms, X or Z | 0.29 | 257 ± 27 | 0.03 | 70.8 | -101 ± 88 | | |
| HY2-C-44 | A0106-IOM | IOM-residue | A | SiC | ms, X or Z | 0.36 | 1462 ± 63 | 0.04 | 36.1 | 13 ± 95 | | |
| HY2-C-45 | A0106-IOM | IOM-residue | A | SiC | ms, X or Z | 0.33 | 393 ± 30 | 0.39 | 63.9 | 45 ± 86 | | |
| HY2-C-46 | A0106-IOM | IOM-residue | A | SiC | ms, X or Z | 0.26 | 400 ± 26 | 0.02 | 63.6 | -42 ± 76 | | |
| HY2-C-47 | C0107-IOM | IOM-residue | C | SiC | ms, X or Z | 0.26 | 347 ± 32 | 0.01 | 66.1 | 64 ± 104 | | |
| HY2-C-48 | C0107-IOM | IOM-residue | C | C-anom | graphite | 0.45 | 1406 ± 22 | 0.00 | 37.0 | -86 ± 57 | | |
| HY2-C-49 | C0107-IOM | IOM-residue | C | SiC | ms, X or Z | 0.28 | 218 ± 29 | 0.92 | 73.1 | -277 ± 112 | | |
| HY2-C-50 | C0107-IOM | IOM-residue | C | SiC | ms, X or Z | 0.37 | 286 ± 22 | 0.52 | 69.2 | -259 ± 77 | | |
| HY2-C-51 | C0107-IOM | IOM-residue | C | SiC | ms, X or Z | 0.28 | 212 ± 28 | 0.36 | 73.4 | -18 ± 114 | | |
| HY2-C-52 | C0107-IOM | IOM-residue | C | SiC | ms, X or Z | <0.1 | 464 ± 74 | 0.00 | 60.8 | 59 ± 244 | | |
| HY2-C-53 | C0107-IOM | IOM-residue | C | SiC | ms, X or Z | 0.28 | 138 ± 24 | 0.06 | 78.2 | -154 ± 110 | | |
| HY2-C-54 | C0107-IOM | IOM-residue | C | SiC | ms, X or Z | 0.30 | 765 ± 33 | 0.10 | 50.4 | -123 ± 273 | | |

### c) Orgueil and Ivuna C-anomalous grains

| Name | Sample | Type | Phase | Group | Size (µm)* | $\delta^{13}C$ (‰) | $Si^-/C^-$ | $^{12}C/^{13}C$ | $\delta^{15}N$ (‰) |
|---|---|---|---|---|---|---|---|---|---|
| Org-C-01 | Orgueil | pressed particle | SiC | ms, X or Z | 0.26 | 795 ± 222 | 5.08 | 49.6 | |
| Org-C-02 | Orgueil | pressed particle | C-anom | ambiguous | 0.34 | -412 ± 83 | 0.14 | 151.4 | 123 ± 240 |
| Org-C-03 | Orgueil | pressed particle | SiC | ms, X or Z | 0.32 | 1500 ± 110 | 1.14 | 35.6 | -221 ± 278 |
| Org-C-04 | Orgueil | pressed particle | SiC | ms, X or Z | 0.18 | 2737 ± 335 | 1.56 | 23.8 | 753 ± 1015 |
| Ivuna-C-01 | Ivuna HK3 | thin section | SiC | ms, X or Z | 0.18 | 549 ± 149 | 0.78 | 57.4 | |
| Ivuna-C-02 | Ivuna HK3 | thin section | SiC | ms, X or Z | 0.21 | 2190 ± 338 | 1.09 | 27.9 | |
| Ivuna-C-03 | Ivuna HK3 | thin section | C-anom | ambiguous | 0.15 | 1072 ± 196 | 0.42 | 42.9 | |
| Ivuna-C-04 | Ivuna HK3 | thin section | SiC | ms, X or Z | 0.19 | 900 ± 191 | 0.62 | 46.8 | |
| Ivuna-C-05 | Ivuna HK3 | thin section | SiC | ms, X or Z | 0.17 | 1520 ± 143 | 0.63 | 35.3 | |
| Ivuna-C-06 | Ivuna HK3 | thin section | SiC | ms, X or Z | 0.21 | 919 ± 119 | 0.62 | 46.4 | |
| Ivuna-C-07 | Ivuna HK3 | thin section | SiC | ms, X or Z | 0.28 | 1030 ± 165 | 0.54 | 43.8 | |
| Ivuna-C-08 | Ivuna HK3 | thin section | SiC | ms, X or Z | 0.28 | 596 ± 148 | 0.89 | 55.7 | |
| Ivuna-C-09 | Ivuna HK3 | thin section | C-anom | likely organic | 0.20 | -226 ± 54 | 0.08 | 115.0 | |
| Ivuna-C-10 | Ivuna HK3 | thin section | SiC | AB | 0.30 | 10045 ± 660 | 0.99 | 8.1 | |
| Ivuna-C-11 | Ivuna HK3 | thin section | C-anom | ambiguous | 0.17 | 2084 ± 439 | 0.50 | 28.9 | |
| Ivuna-C-12 | Ivuna HK3 | thin section | C-anom | ambiguous | 0.26 | 1290 ± 238 | 0.25 | 38.9 | |
| Ivuna-C-13 | Ivuna HK3 | thin section | SiC | ms, X or Z | 0.22 | 513 ± 128 | 0.77 | 58.8 | |
| Ivuna-C-14 | Ivuna HK3 | thin section | SiC | AB | 0.30 | 10021 ± 334 | 0.82 | 8.1 | |
| Ivuna-C-15 | Ivuna HK3 | thin section | C-anom | ambiguous | 0.29 | 3484 ± 312 | 0.59 | 19.8 | |
| Ivuna-C-16 | Ivuna HK3 | thin section | SiC | ms or AB | 0.31 | 6375 ± 276 | 0.89 | 12.1 | |
| Ivuna-C-17 | Ivuna HK3 | thin section | SiC | ms, X or Z | 0.16 | 1549 ± 228 | 0.43 | 34.9 | |

Presolar grain group definitions can be found in Zinner (2014, and references therein). ms = mainstream SiC; mtome = microtome; Microtomes were mounted on an Si substrate and Si/C ratios could not be determined.

IOM = insoluble organic matter (cf. Yabuta et al. 2022)

*ROI-diameter. ROI sizes were defined based on the full width at half maximum of the anomalous region. Based on image simulations by Barosch et al. (2022a), a 120 nm beam size correction was applied to grains with apparent diameters of <0.25 µm in the ion image.

**Grains first reported by Yabuta et al. (2022).

$\delta^{13}C = [(^{13}C/^{12}C)_{grain}/(^{13}C/^{12}C)_{std} - 1\,] \times 1000$; $(^{13}C/^{12}C)_{std} = 0.011237$ (VPDB)

$\delta^{15}N = [(^{15}N/^{14}N)_{grain}/(^{15}N/^{14}N)_{std} - 1\,] \times 1000$; $(^{15}N/^{14}N)_{std} = 0.00367$ (atmospheric N)

$\delta^{x}O = [(^{x}O/^{16}O)_{grain}/(^{x}O/^{16}O)_{std} - 1\,] \times 1000$; $x = 17, 18$; $(^{17}O/^{16}O)_{std} = 0.0003829$, $(^{18}O/^{16}O)_{std} = 0.0020052$ (VSMOW)

$\delta^{x}Si = [(^{x}Si/^{28}Si)_{grain}/(^{x}Si/^{28}Si)_{std} - 1\,] \times 1000$; x = 29, 30; $(^{29}Si/^{28}Si)_{std} = 0.050633$; $(^{30}Si/^{28}Si)_{std} = 0.033474$

Table 2: Matrix-normalized presolar grain abundances (ppm) in Ryugu samples and CI chondrites.

| Sample | Sample type | Chamber | Measured area ($\mu m^2$) | O-anomalous grains #grains | O-anomalous grains abun. (ppm) | O-anomalous grains error +/- | SiC grains #grains | SiC grains abun. (ppm) | SiC grains error +/- | Carbonaceous grains #grains | Carbonaceous grains abun. (ppm) | Carbonaceous grains error +/- |
|---|---|---|---|---|---|---|---|---|---|---|---|---|
| A0058-2 | thin section | A | 18408 | 1 | 1.2 | 2.8/1.0 | 9 | 25.3 | 11.6/8.2 | 9 | 20.7 | 9.5/76.8 |
| A0108-13 | pressed particles | A | 4631 | | not measured | | 4 | 17.8 | 14.0/8.5 | 0 | 0.0 | - |
| C0002 matrix | thin section | C | 3674 | 0 | 0.0 | - | 0 | 0.0 | - | 0 | 0.0 | - |
| C0002 clast 1* | thin section | C | 4907 | 2 | 25.4 | 33.5/16.4 | 4 | 47.4 | 37.5/22.7 | 1 | 7.0 | 16.2/5.8 |
| C0002 clast 2* | thin section | C | 3695 | 0 | 0.0 | - | 3 | 11.5 | 11.2/6.2 | 1 | 8.9 | 20.5/7.4 |
| C0109-2 | pressed particles | C | 3900 | | not measured | | 3 | 37.3 | 36.3/20.3 | 0 | 0.0 | - |
| Ivuna | thin section | | 23110 | 0 | 0.0 | - | 12 | 23.6 | 9.0/6.7 | 5 | 8.1 | 5.5/3.5 |
| Orgueil | pressed particles | | 8054 | | not measured | | 3 | 19.7 | 19.2/10.7 | 1 | 11.5 | 26.5/9.5 |
| (combined)** | Ryugu A | | 23039 | 1 | 1.2 | 2.8/1.0 | 13 | 23.8 | 8.6/6.5 | 9 | 16.6 | 7.6/5.4 |
| | Ryugu C | | 16176 | 2 | 10.1 | 13.4/6.6 | 10 | 26.0 | 11.1/8.1 | 2 | 4.2 | 5.5/2.7 |
| | Ryugu total | | 39215 | 3 | 4.8 | 4.7/2.6 | 23 | 24.7 | 6.3/5.1 | 11 | 11.4 | 4.6/3.4 |
| | CI chondrites | | 31164 | 0 | 0.0 | - | 15 | 22.6 | 7.5/5.8 | 6 | 9.0 | 5.4/3.6 |

*Clast 1+2 are less altered lithologies in section C0002 (Fig. 1b; Kawasaki et al. 2022).

**Total area measured for C-anomalous grains in Ryugu; areas analyzed for O-anomalous grains are slightly smaller as no O isotopes were measured in the pressed particles.

Errors are 1σ (Gehrels 1986).

Abun. = abundances

**Figures**

Fig. 1:

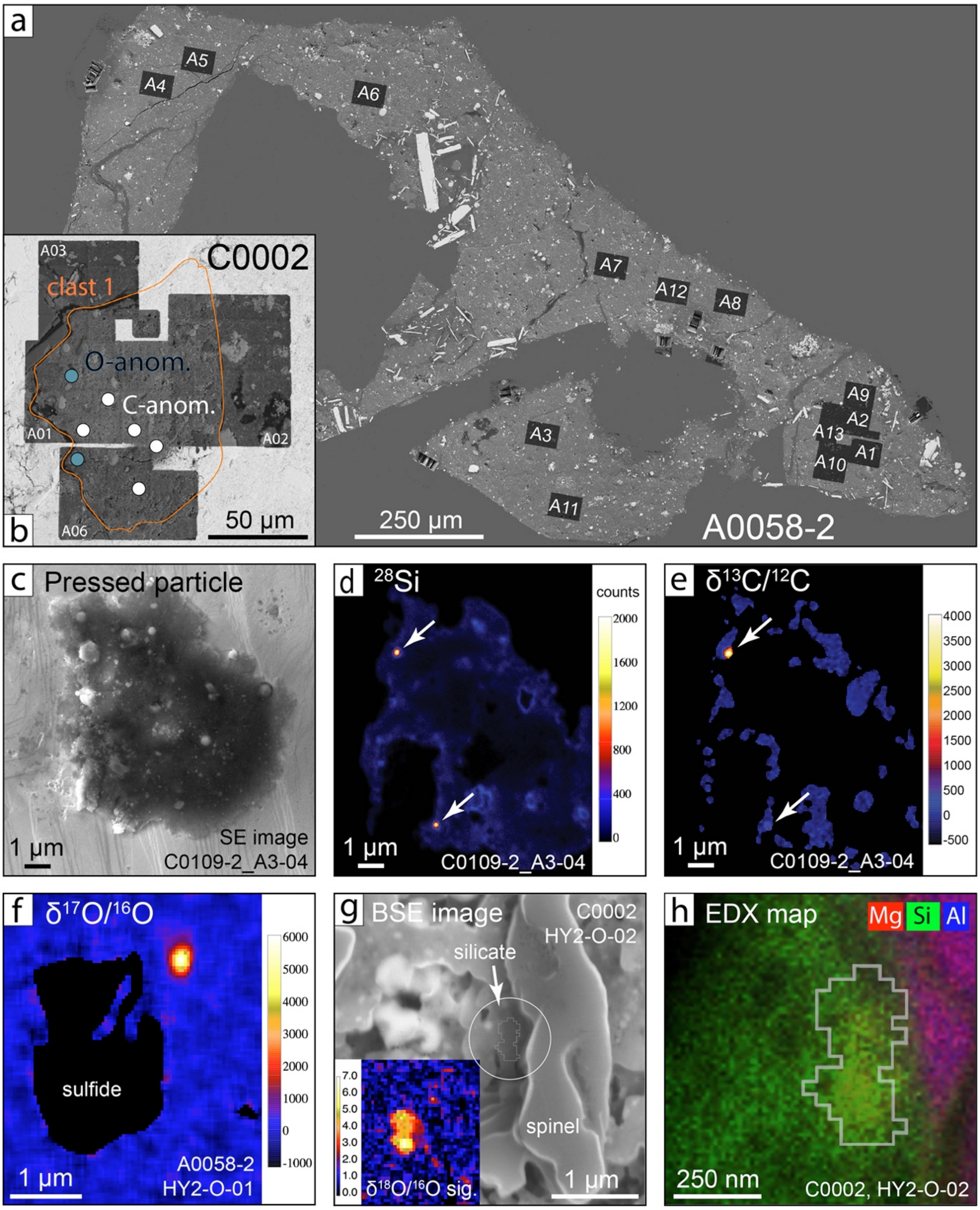

a) Backscattered electron (BSE) image of Ryugu thin section A0058-2. Every black area consists of ~20 NanoSIMS maps measured. b) An area in section C0002 with a less-altered lithology than the surrounding Ryugu matrix ("clast 1"; BSE image). This area contains Mg-rich olivine, low-Ca pyroxenes and spinel grains with sizes up to ~15 µm (Kawasaki et al. 2022). Two of three O-anomalous grains identified in Ryugu, including one likely presolar silicate (g–h), were found in this region. c–e) Secondary electron (SE) image of a Ryugu particle pressed into gold foil in which two presolar SiC grains were detected. The C-anomalous regions, indicated by the white arrows, are clearly associated with $^{28}$Si hotspots. f) $^{17}$O-rich presolar oxide found in Ryugu A0058-2 matrix. g–h) This O-anomalous presolar grain was found in the less-altered area shown in (b). The inlet in (g) shows a $\delta^{18}$O sigma image in which every pixel represents the number of standard deviations from the average values. The grain is probably a presolar silicate as Si is present in the EDX map, and Al was neither detected in the EDX map nor the NanoSIMS ion image, unlike the adjacent spinel ($MgAl_2O_4$), purple in color in (h).

Fig. 2:

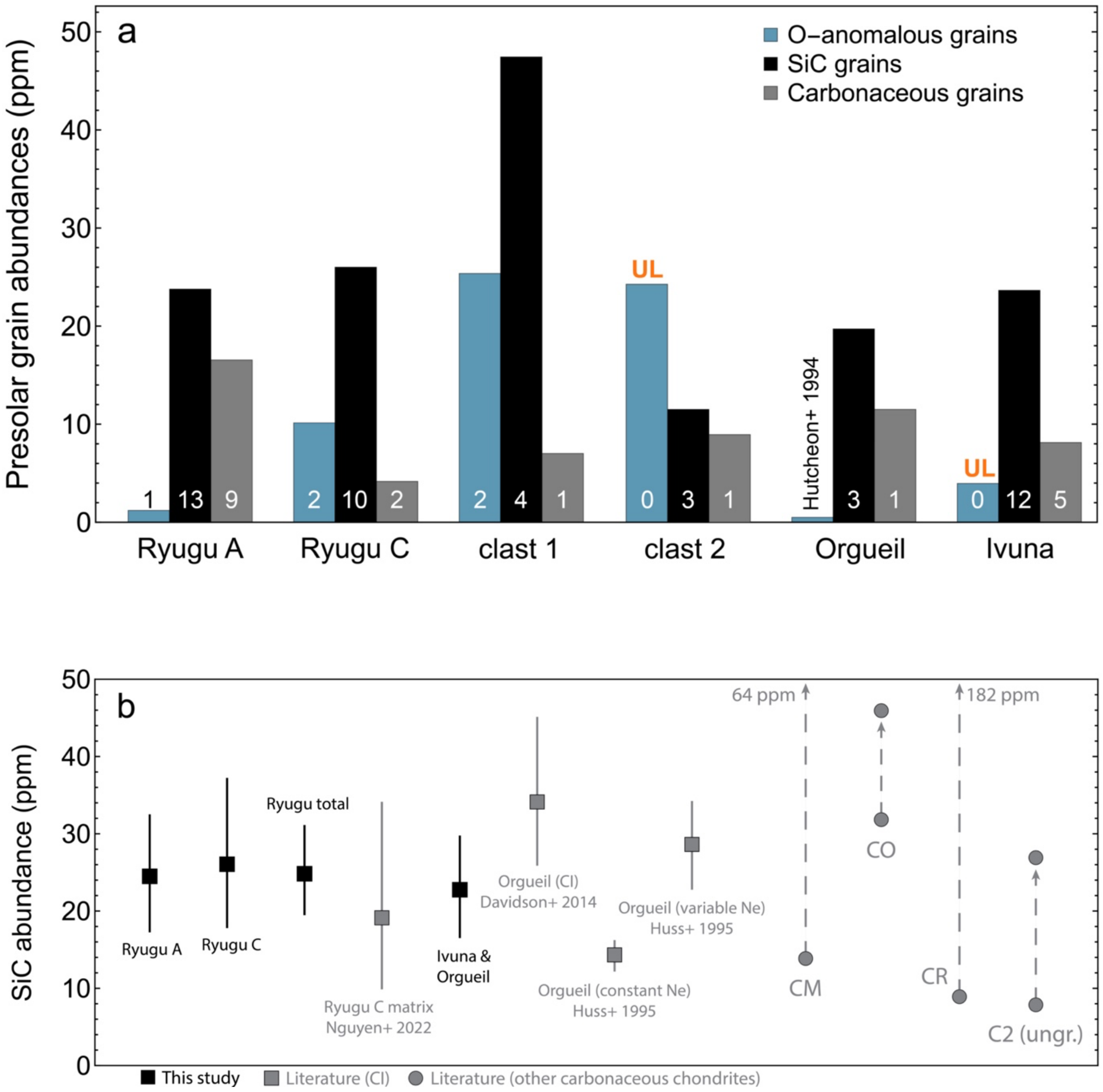

a) Measured matrix-normalized abundances of presolar O-anomalous, SiC and C-anomalous carbonaceous grains in Ryugu samples and CI chondrites (in ppm). The numbers on the bars indicate the number of grains detected. Ryugu A and C are the total abundances measured in chamber A and C samples (including abundances from clast 1 and 2). Clast 1 and 2 are presolar grain abundances in less altered lithologies in section C0002 (cf. Fig. 1b). No O-anomalous grains were detected in Ivuna. O-anomalous grains were not analyzed in Orgueil and the displayed abundance was taken from Hutcheon et al. (1994) and based on a single oxide. UL: Estimated 1σ upper limit for non-detection (Gehrels 1986). Uncertainties are 1σ and listed in Table 2.

b) Matrix-normalized presolar SiC abundances (in ppm) in Ryugu and CI chondrites are compared to literature data (gray). The Orgueil SiC abundances by Davidson et al. (2014) were determined in acid-resistant organic residues, and the abundances reported by Huss & Lewis (1995) were estimated from noble gas analyses (constant and variable Ne-E(H)). All uncertainties are 1σ (Gehrels 1986). Uncertainties for the noble gas analyses by Huss & Lewis (1995) were taken from Table 2 in Davidson et al. (2014). For the other carbonaceous chondrites, lowest to highest published NanoSIMS-based SiC abundances are displayed. CM: Murchison (Davidson et al. 2014) – Jbilet Winselwan (Leitner et al. 2020); CO: Dominion Range 08006 (combined data from Haenecour et al. 2018 and Nittler et al. 2018b) – Allan Hills 77307 (combined data from Davidson et al. 2014 and Haenecour et al. 2018); CR: Isheyevo (Leitner et al. 2018) – Grove Mountains 021710 (Zhao et al. 2013); C2-ungr.: Adelaide (Floss & Stadermann 2012) – Acfer 094 (Davidson et al. 2014).

Fig. 3:

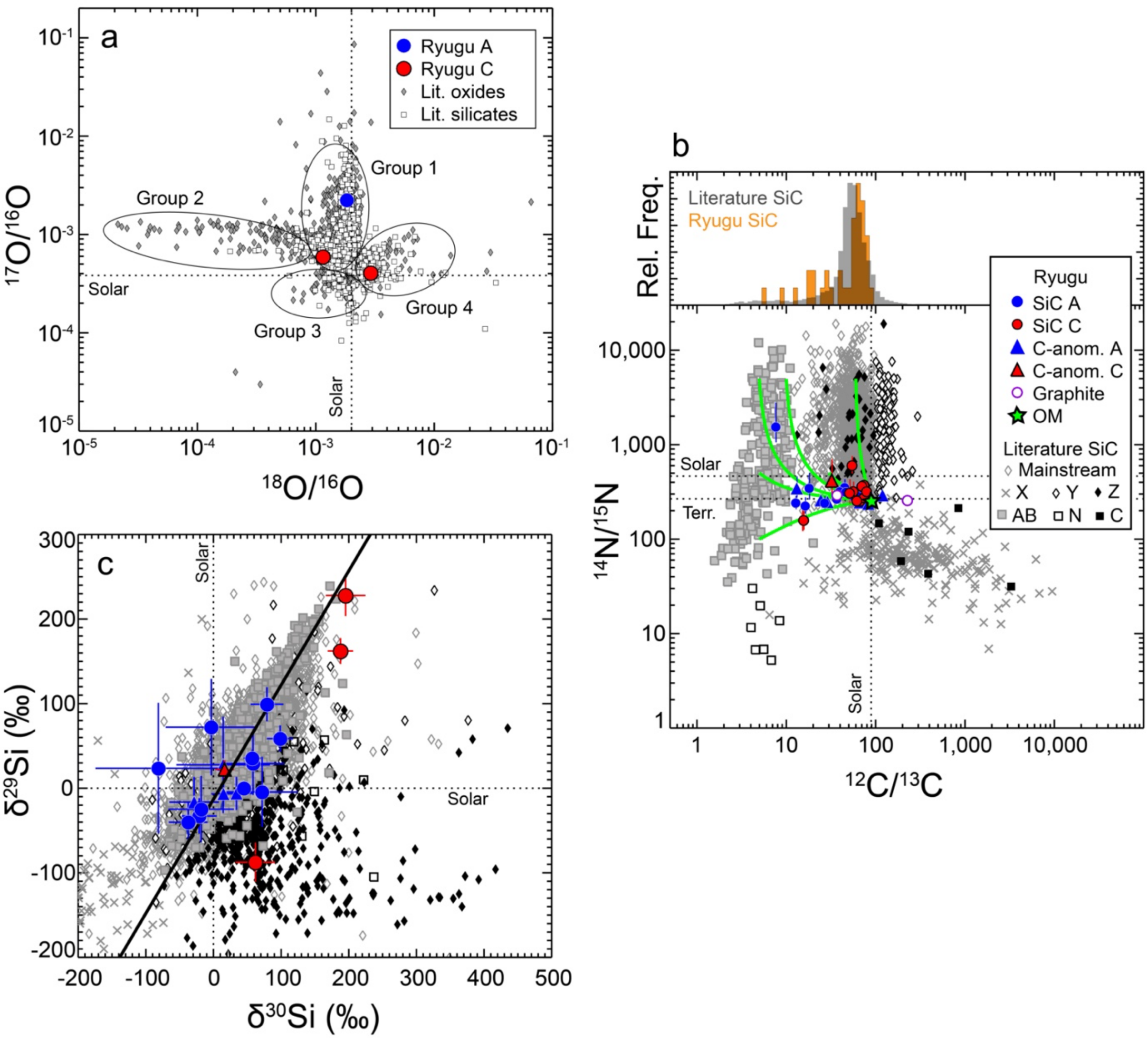

a) Oxygen three-isotope plot of two presolar oxides and a silicate grain in Ryugu. The literature data are taken from a large number of sources (see Floss & Haenecour 2016; Nittler et al. 2021, and references therein). b) C and N isotopic compositions of several SiC and carbonaceous presolar grains in Ryugu compared to literature data (taken from the presolar grain database; Stephan et al. 2020, and references therein). Green curves indicate mixing curves between select compositions and average Ryugu organic matter (OM; Yabuta et al. 2022), indicating that Ryugu presolar grain compositions, particularly N isotopes, have been somewhat contaminated by the ubiquitously present organic matter. c) Si isotopic ratios of Ryugu presolar grains are compared to literature data for meteoritic SiC (cf. presolar grain database; Stephan et al. 2020). The solid line is the best-fit line to mainstream SiC (Zinner et al. 2007).